\documentclass[10pt,twocolumn,english,preprint,prl,amsfonts,amssymb,byrevtex,tightenlines,nobalancelastpage]{revtex4}
\usepackage[T1]{fontenc}
\usepackage[latin9]{inputenc}
\usepackage{amsmath}
\usepackage{graphicx}
\usepackage{amssymb}

\makeatletter

\DeclareRobustCommand{\greektext}{%
  \fontencoding{LGR}\selectfont\def\encodingdefault{LGR}}
\DeclareRobustCommand{\textgreek}[1]{\leavevmode{\greektext #1}}
\DeclareFontEncoding{LGR}{}{}

\newcommand{\lyxmathsym}[1]{\ifmmode\begingroup\def\b@ld{bold}
  \text{\ifx\math@version\b@ld\bfseries\fi#1}\endgroup\else#1\fi}

\@ifundefined{textcolor}{}
{%
 \definecolor{BLACK}{gray}{0}
 \definecolor{WHITE}{gray}{1}
 \definecolor{RED}{rgb}{1,0,0}
 \definecolor{GREEN}{rgb}{0,1,0}
 \definecolor{BLUE}{rgb}{0,0,1}
 \definecolor{CYAN}{cmyk}{1,0,0,0}
 \definecolor{MAGENTA}{cmyk}{0,1,0,0}
 \definecolor{YELLOW}{cmyk}{0,0,1,0}
 }

\usepackage{amsfonts}
\providecommand{\U}[1]{\protect\rule{.1in}{.1in}}

\makeatother

\usepackage{babel}

\makeatother

\usepackage{babel}

\begin{document}

\title{Indirect Magnetic-Field-Tuned Superconductor-Insulator Transitions
and Weak Localization of Bosons of Quasi-Two Dimensional Metal Films}

\author{Yen-Hsiang Lin and A. M. Goldman}

\affiliation{School of Physics and Astronomy, University of Minnesota, 116 Church
St. SE, Minneapolis, MN 55455, USA}

\pacs{PACS number}
\begin{abstract}
Magnetic field and electrostatically tuned superconductor-insulator
(SI) transitions of ultrathin metal films with levels of disorder
that place them near the disorder-tuned SI transition appear to be
direct, continuous quantum phase transitions. When films with lower
levels of disorder are subjected to a perpendicular magnetic field,
instead of a direct transition, a mixed superconductor-nonsuperconductor
regime emerges at the lowest temperatures. The zero temperature limit
of the resistance is either insulating or superconducting, depending
upon the value of the field, suggesting that the behavior in this
limit is governed by percolation physics. At high fields and low temperatures,
in the nominally insulating regime, the resistance rather than the
conductance is found to be a logarithmic function of temperature corresponding
to predicitons for the weak localization of bosons. 
\end{abstract}

\date{10/04/09}

\maketitle
Highly-disordered, homogeneous, quench-condensed, ultrathin films
of metals can be tuned between superconducting and insulating behavior
by magnetic field, thickness or carrier density \cite{Markovic,Parendo1}.
These transitions are of interest because they are among the simplest
of quantum phase transitions, and the field and carrier density tuned
transitions appear to belong to the (2+1)D XY universlity class. Recently
there has been increased interest in highly-disordered, amorphous
compounds such as In$_{2}$O$_{3}$ and polycrystalline TiN, ranging
in thickness from tens to hundreds of Angstroms. In addition to undergoing
superconductor-insulator (SI) transitions tuned by perpendicular field,
these systems exhibit magnetoresistance, $R(B)$, peaks in the insulating
regime\cite{Gant,Samband1,Steiner2,Baturina,Steiner1}. Although this
effect was first observed and interpreted as evidence of a Bose insulating
regime by Paalanen, Hebard, and Ruel \cite{Palaanen} almost two decades
ago, recent experiments have displayed enhancements of resistance
by many more orders of magnitude\cite{Samband2,Vinokur}. Furthermore,
for less disordered films, an intermediate metallic phase has been
reported to occur between the superconducting and insulating regimes\cite{Steiner3}.
What is surprising is that in work on the much thinner, atomically
disordered (amorphous), quench-condensed films, peaks in $R(B)$ in
the insulating regime of have only been observed in films patterned
with a nanohoneycomb array of holes\cite{Nguyen}, and there appears
to not be an instrinsic intermediate metallic regime. In most of the
measurements, normal resistances were close to the critical values
associated with thickness-tuned SI transition. The present study,
which was motivated by a search for magnetoresistance peaks in quench-condensed
films, extends the regime of parameter space to normal resistances
below criticality, yet not so far below that the field-tuned transition
is to a metallic state.

There have been two striking findings. The first is the behavior in
the high-field limit, where the resistance, $R(T)$, rather than the
conductance, $G(T),$ was found to vary as $lnT$. This is suggestive
of the \emph{weak} boson localization phenomenon predicted by Das
and Doniach \cite{Das}, and reported for under-doped cuprates whose
superconductivity was completely quenched by magnetic fields \cite{Ando}.
In addition there was no peak in $R(B)$ in fields up to 10 T. A second
finding is that at high temperatures (around $2K\sim4K$) the data
could be collapsed using a finite size scaling analysis. With decreasing
temperature scaling broke down because curves of $R(T)$ became nonmonotonic
functions of temperature in the manner suggestive of thickness tuned
SI transitions of granular films \cite{Parendo2}. This change in
the physics with decreasing temperature is reminiscent of the approach
to a quantum critical point of some stongly correlated electron systems
in a magnetic field, in which new physics turns on before the quantum
critical point is reached \cite{Mathur}. Mixed phases in the case
of the field-driven SI transition have been discussed in the theoretical
literature \cite{Spivak,Dubi}.

The present investigations were carried out using films grown on (100)
SrTiO$_{3}$ (STO) single-crystal substrates. Platinum electrodes,
100\AA{} thick, were deposited \textit{ex situ} onto the substrate's
epi-polished front surfaces to form a configuration suitable for both
resistance and Hall resistance measurements. The substrate was then
placed in a Kelvinox-400 dilution refrigerator/UHV deposition apparatus
\cite{Hernandez}. A 10\AA{} thick under-layer of \textit{a-}Sb and
successive layers of \textit{a-}Bi were thermally deposited \textit{in
situ} under ultra-high vacuum conditions through shadow masks onto
the substrate's front surface. The substrate was held at about 7 K
during the depositions. Films grown in this manner are believed to
be homogeneously disordered on a microscopic, rather than on a mesoscopic
scale \cite{Strongin}. The sample measurement lines were heavily
filtered so as to minimize the electromagnetic noise environment of
the film. To avoid complications arising from this filtering, measurements
were made using DC, rather than AC, methods. The ion and turbo pumps
connected to the growth chamber were electrically isolated from the
cryostat using vacuum nipples with ceramic spacers.

Figure 1 shows a series of curves of $R(T)$ as a function of magnetic
field charting the transition from superconductor to insulator. %
\begin{figure}[ptb]
 \centering{}\includegraphics[width=3.0441in,height=2.9283in]{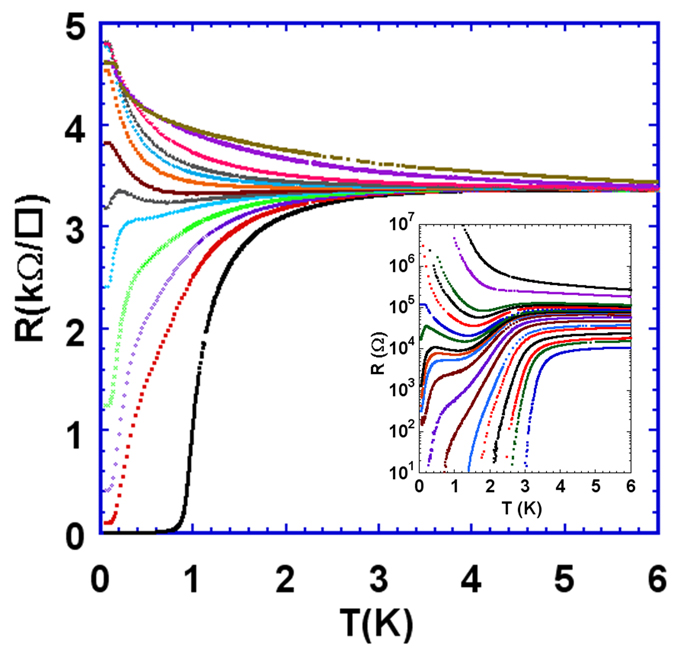}\caption{\emph{a}-Bi (with a 10\AA{}\ thick a-Sb underlayer) film in different
perpendicular magnetic fields. The values of field from top to bottom
are 10, 5, 2.5, 2, 1.8, 1.6, 1.4, 1.3, 1.2, 1, 0.8, 0.6, 0 Tesla.
Inset: Sheet resistance vs. temperature for the thickness-tuned superconductor-insulator
transition of granular (without an underlayer) Bi films adopted from
Ref. 15. The thicknesses are 21.11 (top), 21.41, 21.75, 21.83, 21.94,
22.04, 22.10, 22.14, 22.17, 22.23, 22.38, 22.63, 22.95, 23.37, 23.66,
24.52, 25.40, 26.07, and 27.56 \AA{}(bottom). Notice that these two
plots involve films with different morphologies and different tuning
methods. }

\end{figure}

The significant feature is the non-monotonic behavior of $R(T)$ at
values of magnetic field that are close to those separating superconducting
and nonsuperconducting behavior. This strongly suggests that there
is a range of magnetic field over which the system breaks into a mixture
of superconducting and nonsuperconducting regions at nonzero temperatures
despite the homogeneity of the film. Behavior of this sort is usually
found for granular films. To make the point, the inset of Fig. 1 shows
the evolution from insulator to superconductor as a function of thickness
for a granular film, which also shows a non-monotonic variation of
$R(T)$ \cite{Parendo2}, but with much larger scale in both resistance
and temperature.

This 11.2\AA{} thick film is homogeneous even though it exhibits non-monotonic
behavior at nonzero temperature within a range of fields over which
it is undergoing a transition from superconducting to insulating behavior
in the low temerature limit. Hall effect measurements support this.
They reveal the areal carrier concentration to $1.5\times10^{16}$
cm$^{-2}$, which is close the value expected for a metal. From the
carrier concentration, one can obtain the Fermi wavevector $k_{F}=$$2.5\times10^{-8}$
cm$^{-1}$. Combining this with result for the sheet resistance, which
gives $k_{F}\ell\backsim6$, we find the electronic mean free path
$\ell$ to be around 2.4\AA{}. This indicates that the disorder is
on an atomic rather than a mesoscopic scale. The carrier concentration
would not be expected to be as high in a granular film.

We now turn to the details of the transport. The conductance $G$
can be fit by $lnT$ in the normal state in zero field and in low
magnetic fields. The coefficient of $lnT$ is $5.23\times10^{-6}$
$\Omega$$^{-1}\square$ which agrees with the theory of weak localization
including electron-electron interactions, for the case of strong spin-orbit
coupling \cite{Abrahams}. The coefficient of the logarithm also agrees
with that found in previous work \cite{Parendo1}. A sample fit, with
the range of applicability clearly delineated is shown in Fig. 2a.%
\begin{figure}[ptb]
 \centering{}\includegraphics[width=2.7207in,height=4.9in]{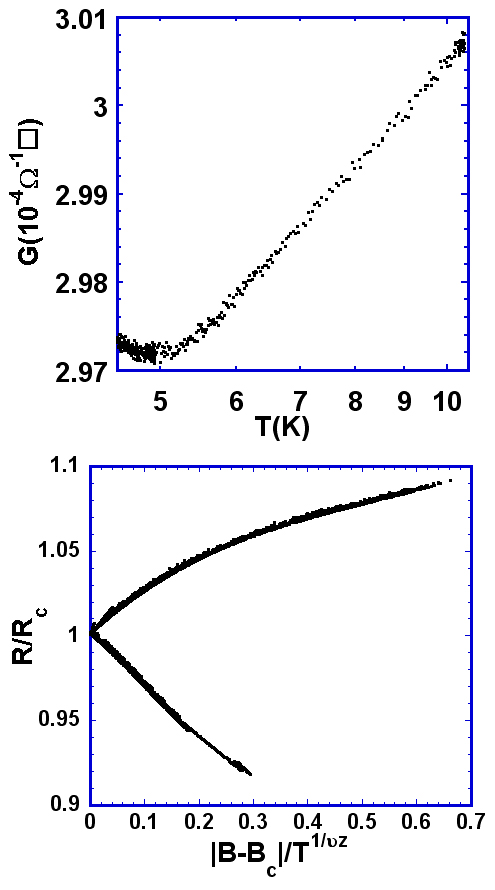}\caption{$G(T)$ in zero magnetic field. (b) Scaling analysis for the perpendicular
magnetic field tuned transistion between $2K<T<4K$ and $0.6T<B<5T$.
Resistance, normalized to the critical resistance (3384$\Omega$),
is plotted as a function of $|B-B_{c}|/T^{1/\upsilon z}$ with $\upsilon z$=0.4
and $B_{c}$=1.58T.}

\end{figure}

In the vicinity of the crossover magnetic field, the resistance at
relatively high temperatures could be loosely fit with the form $R\sim Rexp(T_{0}/T)$,
with a negative value of $T_{0}$ for films that smoothly become superconducting
and a positive value for those which ultimately become insulators.
This suggested that $R(T,B)$ might be collapsed using the finite-size
scaling relation suggested by Fisher \cite{Fisher}. Figure 2b shows
the result of this analysis for $2K<T<4K$ and $0.6T<B<5T$ employing
the form $R/R_{c}\sim F\left(|B-B_{c}|/T^{1/\nu z}\right)$ with the
critical exponent product $\lyxmathsym{\textgreek{n}}z=0.4$. The
scaling fails when applied to data outside of these ranges.

Well into the insulating regime, where one might expect a return to
$G\backsim lnT$, but one finds instead, that $R$ rather than $G$
is better described by $lnT$. A comparison of these two forms for
data obtained at a field of 10T is presented in Fig 3. This relationship
was also found at lower values of magnetic field. We compared the
values of chi square $\chi^{2}=\sum\left[(R_{i}-R_{fit}(T_{i}))/\sigma_{i}\right]$
for the two functional forms ($R\backsim lnT$ and $G\backsim lnT$),
where the $R_{i}$ are the data points, $R_{fit}(T_{i})$ is the fitted
function, and the $\sigma_{i}$ are the errors of the data points.
We include data over the range from 100 mK to 1 K, and in magnetic
fields of 2.5, 3, 4 and 5 T. For $R\backsim lnT$ the values were
6104, 7063, 2932, and 1291 respectively, whereas for $G\backsim lnT$
they were 37276, 40298, 19957, and 8216, respectively. This makes
the case for $R\backsim lnT$ being the better description of the
data.

Futhermore, if one forces the data to be fit by the functional form
$G\backsim lnT$, as shown in Fig 3(b), the coefficient of $lnT$
is nearly three times larger than that found in the high-temperature
low-field regime. For example, in zero field in the temperature range
of $5K\sim10K$, the coefficient is $5.23\times10^{-6}\Omega^{-1}\square$,
whereas it is $1.63\times10^{-5}\Omega^{-1}\square$ at $10T$ over
the temperature range from $0.1K\thicksim1K$ when forced. This indicates
that the resistance of the film in the high field, low temperature
regime increases more rapidly than that of a quantum corrected metal.
\begin{figure}[ptb]
 \centering{}\includegraphics[width=2.72in,height=4.9in]{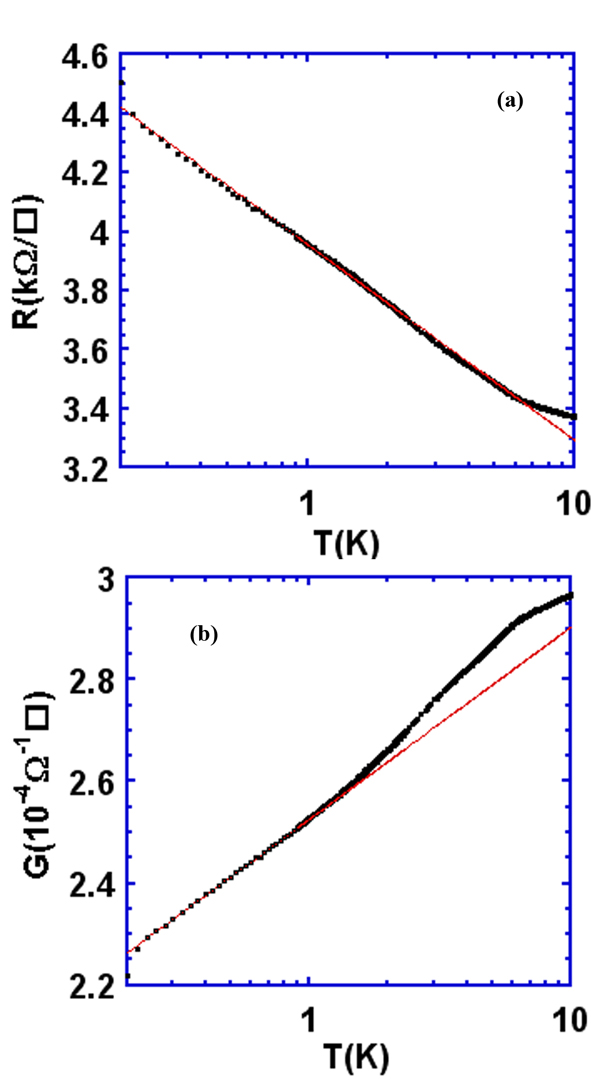}\caption{$logT$ with the results of fit in a 10 T field. The red thin lines
are the fitting lines. The resistances are fit better by $lnT$ over
a larger temperature range (up to 6K) than the conductances.}

\end{figure}

Figure 4 provides a global representation of the regimes over which
$R(T)$ can be described by various forms as well as crossover and
nonmonotonic regimes which cannot be described by simple functional
forms. The data clearly suggest that the evolution from a superconducting
ground state to a nonsuperconducting state is indirect, involving
an intermediate regime at nonzero temperature in which there are superconducting
islands in a nonsuperconducting matrix\cite{Spivak}\cite{Dubi}.
\begin{figure}[ptb]
 \centering{}\includegraphics[width=3.0415in,height=3.0512in]{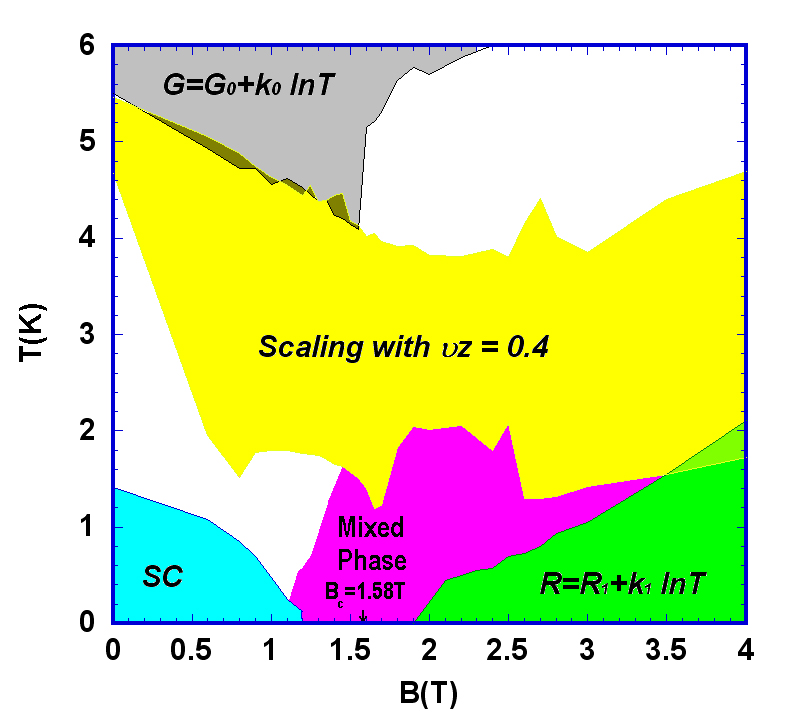}\caption{$R\sim lnT$ and $G\sim lnT$ regimes correspond to $0.1\%$ difference
between data points and the results of the fits. The boundaries of
the scaling regime were set using the same criterion. The indicated
value of $B_{c}$ corresponds to the critical field of the scaling
analysis.}

\end{figure}

In the case of quench-condensed, disordered superconducting films
with disorder levels closer to the disorder- or electric field-tuned
transitions than the present films, in the high-field regime the behavior
is clearly insulating although there is no magnetoresistance peak
which has been taken as evidence that the insulator is a Cooper pair
insulator. The latter behavior has only been observed in compound
amorphous films or in quench condensed films patterned with nanohoneycomb
hole arrays. The high magnetic field regimes of the films of the present
work are not a quantum corrected 2D metals, because as demonstrated
above, $R(T)$ rather than $G(T)$ is better fit by $lnT$ over an
extended range of temperature. The temperature dependence is the same
as that predicted in the work of Das and Doniach \cite{Das}, which
describes a Boson weak localization model. Although the coefficient
of the $lnT$ term is an order of magnitude smaller than that which
they predict, the model may be relevant. Similar behavior has been
observed in underdoped cuprate superconductors when superconductivity
is quenched by a magnetic field \cite{Ando}. Moreover, the coefficient
of $lnT$ in the present work is close to that found in that work.
If one converts the sheet resistance into resistivity, the coefficient
of $lnT$ in a field of 2T is $-6.9\times10^{-5}$ $\Omega$-cm, while
for the case of x=0.13 of La$_{2-x}$Sr$_{x}$CuO$_{4}$, it was found
to be $-7.2\times10^{-5}$$\Omega$-cm when current is in the ab plane.

The fact that at higher temperatures the data can be scaled, suggests
the existence of a quantum critical point. However, a direct quantum
phase transition never occurs, because the two-phase regime develops
as one lowers the temperature. In the limit of zero temperature the
film is either insulating or superconducting so that it is likely
that percolation physics plays a role in the SI transition. The value
of the critical exponent product $\lyxmathsym{\textgreek{n}}z=0.4$
found in the regime in which the data scales is not associated with
any particular model relevant to ultrathin films. However this value
has been found in a Monte Carlo simulation of (2+1)D XY Josephson
junction arrays without disorder and with frustration $f=1/4$ \cite{Lee}.
Furthermore, this simulation exhibits a first order phase transition
when the frustration drops to $f=1/5$. A two-phase regime such as
inferred from our data could also imply a first order quantum phase
transition. Modeling a disordered film with a junction array would
appear to require a Monte Carlo simulation including disorder, which
is not included in the simulation of Lee and Cha. On the other hand,
Kim and Stroud performed a simulation including disorder but without
magnetic field \cite{Kim}. A simulation with disorder and frustration
would be needed to model films such as those reported in this work.

The present results deepen the mystery associated with the superconductor-insulator
transition. Although various models may account for the behavior of
specific systems, there is no model that explains their different
behaviors. For example, it is not clear as to why some transitions
are direct, and with small changes in the level of disorder an intermediate
two-phase regime emerges. Also the reasons for the differences between
the high field regimes of different materials are not known.

This work was supported in part by the National Science Foundation
under grants NSF/DMR-0455121 and NSF/DMR-0854752.

\end{document}